\documentclass{article}
\pdfoutput=1
\usepackage{amsmath}
\usepackage{enumerate}
\usepackage{amssymb}
\usepackage{graphicx}
\usepackage{multirow}
\usepackage{xcolor}
\usepackage{bm}
\usepackage[normalem]{ulem}

\usepackage{jcappub}

\newcommand{\sumnu}{\Sigma}

\def\m@th{\mathsurround=0pt }
\def\eqalign#1{\null\,\vcenter{\openup1\jot \m@th
 \ialign{\strut\hfil$\displaystyle{##}$&$\displaystyle{{}##}$\hfil
 \crcr#1\crcr}}\,}

\begin{document}

\title{Strong Bayesian Evidence for the  \\ Normal Neutrino Hierarchy}

\author[1]{Fergus Simpson,}
\author[1,2]{Raul Jimenez,}
\author[3,4]{Carlos Pena-Garay,}
\author[1,2]{Licia Verde}

\affiliation[1]{ICCUB, University of Barcelona (UB-IEEC), Marti i Franques 1, Barcelona, 08028, Spain.}
\affiliation[2]{ICREA, Pg. Lluis Companys 23, Barcelona, 08010, Spain.} 
\affiliation[3]{I2SysBio, CSIC-UVEG, P.O.  22085, Valencia, 46071, Spain.}
\affiliation[4]{LSC, Estaci\'on de Canfranc, 22880, Spain.}

\emailAdd{fergus2@gmail.com; raul.jimenez@icc.ub.edu; liciaverde@icc.ub.edu; penagaray@gmail.com;}

\abstract{ The configuration of the three neutrino masses can take two forms, known as the normal and inverted hierarchies. We compute the Bayesian evidence associated with these two hierarchies. Previous studies found a mild preference for the normal hierarchy, and this was driven by the asymmetric manner in which cosmological data has confined the available parameter space. Here we identify the presence of a second asymmetry, which is imposed by data from neutrino oscillations. By combining constraints on the squared-mass splittings \cite{Gonzalez-Garcia:2014bfa} with the limit on the sum of neutrino masses of  $\Sigma m_\nu < 0.13$ eV \cite{2016Cuesta}, and using a minimally informative prior on the masses, we infer odds of 42:1 in favour of the normal hierarchy, which is classified as ``strong" in the Jeffreys' scale.   We explore how these odds may evolve in light of higher precision cosmological data, and discuss the implications of this finding with regards to the nature of neutrinos. Finally the individual masses are inferred  to be $m_1 = 3.80^{+26.2}_{-3.73} \, \text{meV}; m_2 =  8.8^{+18}_{-1.2} \,  \text{meV}; m_3 =  50.4^{+5.8}_{-1.2} \,  \text{meV}$ ($95\%$ credible intervals).}
 
\maketitle

\section{Introduction}

Cosmological surveys provide crucial information on the absolute masses of neutrinos, information that is not accessible  even to future laboratory experiments \cite{Osipowicz:2001sq,Monreal:2009za,Doe:2013jfe}.  Recent analyses have already established a robust upper limit on the  sum of neutrino masses of $\sumnu < 0.13$ eV \cite{2016Cuesta} and   $\sumnu < 0.12$ \cite{Palanque-Delabrouille:2015pga} at 95\% confidence (throughout this work we shall denote $\Sigma m_\nu$ as $\sumnu$ for brevity). This limit is tantalisingly close to the lower limit for the sum of the masses in the inverted hierarchy $\sumnu=0.0982 \pm  0.0010$ eV (68\% C.L.) \cite{Gonzalez-Garcia:2014bfa}.  Because the volume covered by cosmological  surveys keeps increasing, their statistical errors shrink,   soon leading to either (a) a measure the total neutrino mass or (b) the exclusion of the inverted hierarchy, leaving the normal one as the only viable option. It is therefore possible to determine the neutrino hierarchy, even in the absence of a measurement of the total mass e.g., Ref.~\cite{Carbone:2010ik, Audren:2012vy,Hamann:2012fe}. Future surveys may even yield a measurement of the mass differences between the lightest and heaviest eigenstates  \cite{Jimenez:2010ev}. This measurement would also provide a powerful consistency check: the measured mass splitting  from cosmology  should be consistent with the squared mass splitting from oscillations. Given the current knowledge about mass splitting and cosmological constraints on the total mass,  can anything be said about the neutrino mass hierarchy? 

Since current cosmological upper limits are already close to the inverted hierarchy bound, the volume of parameter space available is heavily restricted. As such, it is useful to employ Bayesian evidence methods in order to determine which hierarchy is favoured by the data. While it seems counterintuitive that one can determine a quantity without having actually measured it, consider the following pedagogical example: if there is a thin stick of length $L$ hidden underneath a cloth, one does not necessarily need to reveal the object in order to learn of its orientation.  At the start, it is considered equally likely to be in either a horizontal or vertical position.  If the cloth is gradually withdrawn in the horizontal direction, and no sign of the stick emerges, one need not wait until a residual length less than $L$ is left unexposed to be confident of a vertical orientation.  Even while a full $2L$ of the surface remains covered, the odds are already stacked $2:1$ in favour of a vertical orientation.   

Our main objective is to identify which hierarchy is preferred by the Bayesian evidence ratio, and understand how this is likely to evolve as new cosmological data arrives in the near future. Previous inferences on the neutrino hierarchy have been made by  e.g., Refs.~\cite{2016Hannestad, Mena17,referee_request}. One crucial aspect where our work differs from previous work in the literature is the choice  and the dimensionality of the prior, which we shall explore in detail. We will also outline the consequences and implications of identifying the nature of the neutrino hierarchy. 

\section{Cosmology, oscillations, and statistics}

In this section we provide a brief review of the three disciplines that are central to this work.  

\subsection{Cosmology: observational constraints and degeneracies}
 
Massive neutrinos influence both the expansion history and growth of structure in the Universe. Neutrinos of mass $\lesssim 1$ eV become non-relativistic after the epoch of recombination probed by the Cosmic Microwave Background (CMB), and this mechanism allows massive neutrinos to alter
the relationship connecting the physical matter density $\Omega_m h^2$ to the redshift of matter-radiation equality.  The neutrinos' radiation-like behaviour at
early times modulates the peak positions in the CMB angular power spectrum.  However this effect is somewhat degenerate with other cosmological parameters. Neutrinos are also responsible for dampening  the matter power spectrum below their free-streaming scale; see  e.g., \cite{Lesgourgues:2006nd} for a review.  This effect, which  amounts  to a scale dependent power suppression, is straightforward to compute in linear theory \cite{Bond:1980ha} and in the non-linear regime via numerical simulations e.g.,  \cite{Brandbyge:2009ce, Wagner:2012sw, Bird:2011rb}.

Due to these mechanisms, cosmological observables are predominantly sensitive to the sum of neutrino masses $\sumnu$  e.g., \cite{1998PhRvL..80.5255H,2004PhRvD..70d5016L, 2005PhRvL..95a1302W,Lesgourgues:2006nd, 2006PhRvD..73h3520T, 2007JCAP...07..004H,2008PhRvD..77j3008K}. 
While the effect of neutrino mass on the CMB is related to the physical density of neutrinos, and therefore the mass difference between eigenstates can be neglected, individual neutrino masses, however, can have an effect on the large-scale shape of the matter power spectrum.  Neutrinos of different masses have different transition redshifts from relativistic to non-relativistic behaviour,  and their mass splitting changes the details of the radiation-domination to matter-domination regime.   A precise measurement of the matter power spectrum shape can give information on both the sum of the masses and individual masses (and thus the hierarchy), even if the second effect is much smaller than the first \cite{Jimenez:2010ev}.  

In this work we shall utilise cosmological constraints imposed on the sum of the three neutrino masses, $\sumnu$. The notation we shall use to denote the three neutrino mass eigenstates is $m_1$, $m_2$, and $m_3$. Thus there are only two independent neutrino mass squared differences. In this convention  $m_1<m_2$ and $m_1, m_2$  refer to the smaller square mass difference. Thus  there are two possible hierarchies only. In what is known as the normal  mass hierarchy, $m_1$, $m_2$, and $m_3$ are defined in ascending order, $m_1 \leq m_2  \leq  m_3$. For the case of the inverted hierarchy,  the sequence becomes  $m_3  \leq  m_1  \leq  m_2$. In other words the hierarchy does not strictly corresponds to whether we have two small masses and one large or vice versa, this arises only once the results of the measurements of atmospheric and solar oscillations square mass splitting are in hand; the hierarchy is given by the sign of the square mass splitting involving $m_3$.
 
Significant progress has been made by  cosmological surveys to constrain the total mass of neutrinos.  For example, the latest Planck satellite results show that CMB data alone constrain $\sumnu<0.59$  eV  (95\% C.L.) \cite{Ade:2015xua}, where the upper limit is stronger than $\sim 1$ eV because of the extra constraining power offered by secondary effects, mostly lensing. The inclusion of late-time information about the cosmic expansion history, such as that provided by measurements of the Baryon Acoustic Oscillations (BAO), further breaks cosmological parameters degeneracies yielding a limit  $\sumnu<0.17$ eV (95\% C.L.). 

Including information about the shape of the matter power spectrum further tightens these constraints, even when BAO information is not used. 
In Ref.~\cite{2016Cuesta}  the authors have carefully examined the impact of using different galaxy populations and tracers to estimate the power spectrum of the underlying dark matter and concluded that $\sumnu < 0.13$ eV at 95\% confidence. This constraint tightens to  $ \sumnu < 0.11$ eV  if also the latest measurement of the Hubble constant \cite{Riess2016} are included. Ref.~\cite{2016Cuesta} result is consistent with the previous constraint from a different estimate of the power spectrum of dark matter using the Lyman-$\alpha$ forest \cite{Palanque-Delabrouille:2015pga}, $\sumnu < 0.12$ eV (see also \cite{2016Giusarma, 2016Valentino}).  Since  all these measurements are only upper limits  for which the maxima of the  probability distributions  are indistinguishable from zero, unless otherwise stated, we will assume that the probability distributions for $\sumnu$ obtained from cosmological data are peaked at $\sumnu=0$.

All the above constraints assume a standard cosmological model, where the spatial geometry is flat, dark matter (excluding the neutrino component) is cold,  the dark energy is a cosmological constant,  general relativity is the correct description of gravity on large, cosmological scales,  the initial conditions of the Universe are given by a process like inflation  and thus the primordial  power spectrum is a nearly scale invariant  power law. In addition, it is implicitly assumed that modelling of real world effects such as tracer's bias, non-linearities, etc. is sufficiently accurate as not to induce systematic errors that are larger than the reported  ones. 
Relaxing any of these assumptions would weaken the constraints significantly  e.g., \cite{dePutter:2014hza, Namikawa:2010re, 2014Baldi}. 
 
Detecting the effect of neutrino masses on cosmological structure and resolving the neutrino mass scale is well within the reach of up-coming cosmological surveys (e.g., \cite{JKPGV,Carbone:2010ik,Audren:2012vy} and references therein).
  
\subsection{Neutrino Experiments}
 
Solar, atmospheric, reactor and accelerator neutrino experiments have observed neutrino flavor conversion driven by neutrino masses. 
The experiments that mostly contribute to a precise measurement of the neutrino masses are sensitive to the relative phases acquired by relativistic neutrinos on their way from the production zone to the detector. Therefore the neutrino oscillation observables are sensitive to mass splittings, specifically to differences in the square of their masses. Matter effects allow the sign of the mass splittings to be distinguished, a goal which has been achieved in the small mass squared splitting but not the large one. The degeneracy in this second case distinguishes the two hierarchy mass orderings, normal and inverted.  Identifying the neutrino mass hierarchy is a key objective for future neutrino oscillation experiments, in addition to a determination of the complex phase involved in flavor conversion.  

The absolute neutrino mass is not accessible via flavor conversion experiments, but it can be probed using beta decay. Experiments require sufficient mass and very good energy resolution in order to measure the beta spectrum close to the endpoint, where the neutrino energy is the neutrino mass. The experiments measure the effective mass of the electron, which is the combination of neutrino masses weighted by their content in the electron flavor. The present bound derived from the electron mass is 2.3 eV  ($\Sigma<6.9\,$eV) at 90\% confidence.  The KATRIN experiment is expected to significantly improve upon this bound, reaching approximately 0.2 eV  ($\sumnu<0.6\,$eV) at 90\% \cite{Osipowicz:2001sq}. Meanwhile new proposals \cite{Monreal:2009za} like Project 8 \cite{Doe:2013jfe} aim for yet greater sensitivity.

 \subsection{Bayesian Model Selection}
 
In this work  we shall employ Bayesian model selection in order to quantify the manner in which the combination of existing data,  from both cosmology and neutrino experiments, should influence our belief in the normal or inverted hierarchy.   Each hypothesis possesses the same number of free parameters, facilitating a fair comparison between the two. Furthermore their low dimensionality (three masses) permits an exact evaluation of the Bayesian evidence, a process which is often highly impractical for more complex models.

The Bayesian evidence $p(D|{\cal H})$ (the probability of the data $D$ given the hypothesis ${\cal H}$) is largely ignored during conventional analyses of cosmological parameter estimation, as it merely takes the form of a normalising constant in Bayes theorem

\begin{equation}
p(\alpha | D, {\cal H}) = \frac{ p(D | \alpha, {\cal H})p(\alpha | {\cal H}) }{p(D| {\cal H})} \, .
\end{equation}
where  $p(D | \alpha, {\cal H})$ is the likelihood, $\alpha$ denote the parameters of the model,  and $p(\alpha | {\cal H})$ the prior.
However, when it comes to determining our preference for one hypothesis over another, the evidence plays a central role.  In order to evaluate the evidence it may be expressed in terms of the likelihood   and the prior probability  $\pi=p(\alpha|{\cal H})$ as follows   \begin{equation} \label{eq:evidence}
 p(D|{\cal H})=\int p(D|\alpha,{\cal H}) p(\alpha|{\cal H})  d\alpha\,.
 \end{equation}

The ratio of evidences for two competing hypotheses, ${\cal H}_1$ and ${\cal H}_2$,  is referred to as the Bayes factor, and this represents the key quantity in Bayesian model selection,

\begin{equation} \label{eq:bayesfactor}
K = \frac{ p(D|{\cal H}_1)}{ p(D|{\cal H}_2)} \, .
\end{equation}
Mechanisms which lead us to prefer one model over another can be crudely decomposed into two contributing factors: the peak and the breadth of the integrand given in (\ref{eq:evidence}). These two quantities are known as the maximum likelihood value, and the so-called Occam factor. If one hypothesis is capable of fitting the data better than another,  then it will yield a greater value of $p(D|\alpha,{\cal H})$, and this modifies our belief accordingly. Of greater significance in this work is the second contribution to the evidence,
which relates to the predictivity of the model.   This concept was succinctly summarised by MacKay \cite{mackay2003information} as follows:

\emph{``The Occam factor is equal to [...] the factor by which $\mathcal{H}$'s hypothesis space collapses when the data arrive."}

In other words the Occam factor penalises complex models; these can be models having many parameters or, as in this case when the two models we compare have the same number of parameters,  the complexity of the predictions that the model makes. In this case  the Occam factor  penalises models that have to be fine tuned to fit the data. Thus the Bayes factor balances the quality of the fit versus the model complexity, as such it rewards highly predictive models and penalises models with ``wasted" parameter space (i.e., it favours models which  offer a good fit over  large part of the prior volume). Inevitably the evidence calculation thus depends on the choice of prior and prior ranges. Our prior choice will be discussed below. Therefore, in the context of the present work, if a much greater fraction of one hierarchy's three-dimensional parameter volume is excluded by the data, we should favour the other hierarchy. 
 
 \section{Inference of the Neutrino Hierarchy}
  
The Bayesian evidence for the two neutrino hierarchies has recently been considered  by  Refs.~\cite{2016Hannestad, Mena17, referee_request}. Where our work differs most significantly from earlier studies is that we shall explicitly consider the full three-dimensional parameter space $(m_1, m_2, m_3)$. Combining likelihoods in a lower dimensional space is a suboptimal procedure, and often leads to the likelihood of a model being greatly overestimated. For example,  in a projected two dimensional space, two datasets may appear to possess large overlapping areas of high likelihood. Yet when viewed in their higher dimensional space, they may represent wholly disjoint volumes.

 Here we shall also argue that when dealing with parameters whose uncertainty spans several orders of magnitude, such as the neutrino masses, adopting a flat prior artificially favours the highest order of magnitude and may skew the interpretation of the results. As summarised by MacKay \cite{mackay2003information},  \emph{``Ignorance does not correspond to a uniform probability distribution"}.
 We will explore the impact of adopting a less informative prior, thus relaxing its effect on the determination of the hierarchy. Specifically, we adopt Jeffreys' prior, which is the least informative, as proved by Clarke \& Barron \cite{clarke1994jeffreys}.

A further appealing feature of Jeffreys' prior is that it is invariant under reparameterisation. Should one's prior belief of the neutrino mass depend on the units which one uses to measure it, whether we use eV or kg?  If our state of knowledge is the same in each case, then we ought to assign equivalent probabililties to each.  If we are to achieve this status, our prior should be invariant under reparameterisation \cite{mackay2003information}.

We wish to determine appropriate betting odds for which hierarchy offers the correct description of the truth. Therefore we must first establish a prior which describes our state of belief before we have gained any knowledge about the masses or their splittings.   At that stage, before the measurement of neutrino oscillations and thus before having information about the (squared) mass splitting, each of the neutrino masses has an uncertainty that spans many orders of magnitude. A logarithmic prior on the masses naturally incorporates this uncertainty, and is known to be the least informative prior \cite{clarke1994jeffreys}.  Below we discuss how these initial beliefs change once we acquire information on the squared mass splittings.

 \begin{figure}
 \centering
\includegraphics[width=115mm]{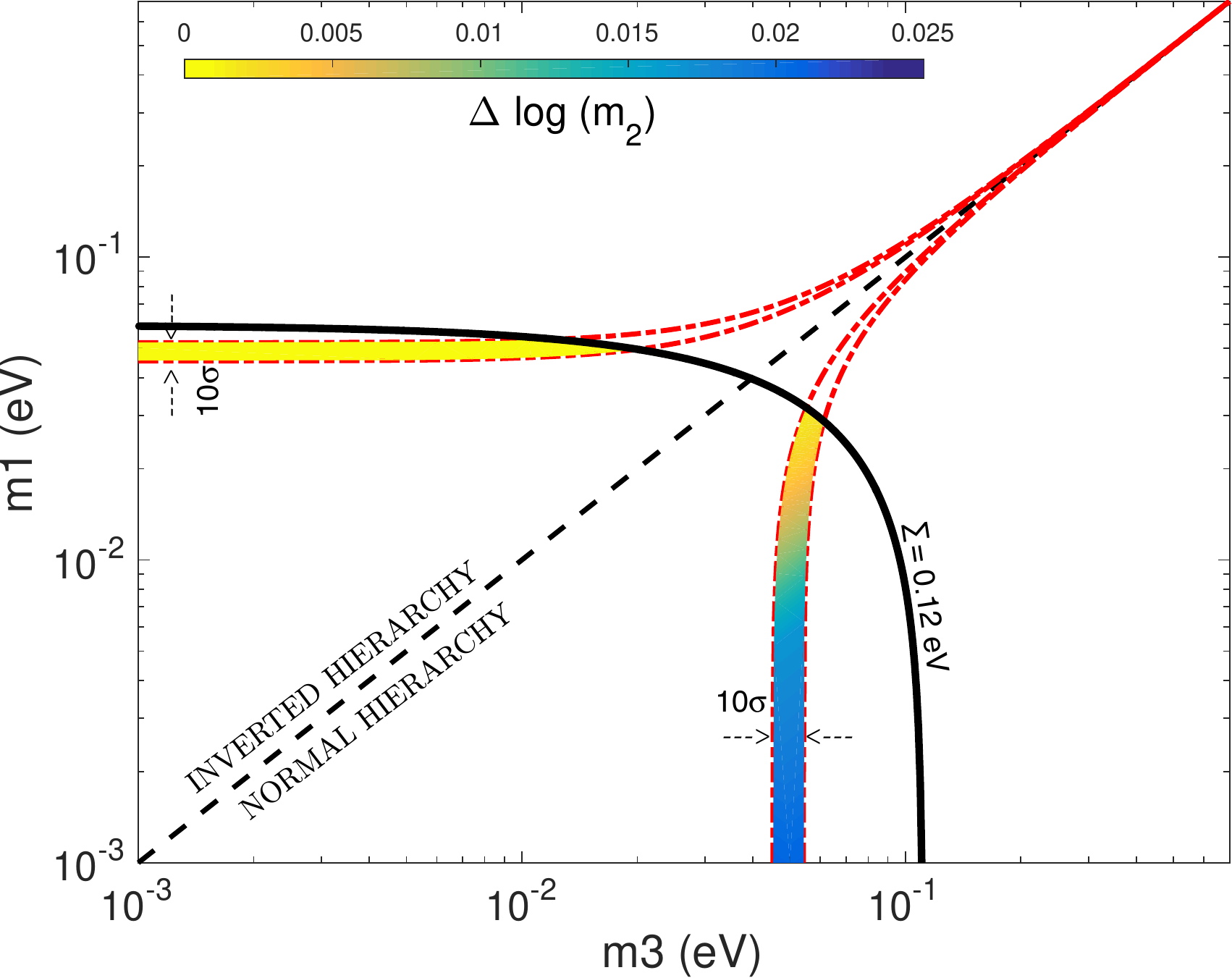}
\caption{A visualisation of the heavily reduced parameter space available in the case of the inverted neutrino mass hierarchy, relative to the normal case.   The red dash-dot contours illustrate constraints on the mass splittings, as imposed by neutrino oscillation experiments (broadened to show $10\sigma$ uncertainties for visualisation purposes). The solid black line corresponds to the combination of a cosmological upper bound on the sum of the neutrino masses $\sumnu < 0.12\, \text{eV}$ with the measurement of $\Delta m_{12}^2$. The diagonal dashed line demarcates the two hierarchies. The colouring of the shaded areas represents the amount of  parameter space available in the third dimension, $\Delta (\log m_2)$.}   \label{fig:m1m3}
\end{figure}

\subsection{Data from Neutrino  Oscillations}

We shall make use of the following squared-mass splittings, as derived from observations of neutrino oscillations using NuFIT v3.0 (2016) \cite{Gonzalez-Garcia:2014bfa}

\begin{equation}
  \label{eq:m12}
m_2^2 - m_1^2 = 7.50  \, (\pm 0.18) \times 10^{-5} \mathrm{eV}^2 \, , 
\end{equation}

\begin{equation}
 \label{eq:m13}
   m_3^2 - m_\ell^2 = 
\begin{cases}
  \phantom{-}   2.524  \, (\pm 0.040) \times 10^{-3} \mathrm{eV}^2   ,& \text{(NH) }  \\
     -2.514 \, (\pm 0.040) \times 10^{-3} \mathrm{eV}^2       ,& \text{(IH)}
\end{cases}
\end{equation}
where $m_\ell$ denotes  $m_1$ and $m_2$ for the normal (NH) and inverted (IH) hierarchies respectively, and we shall approximate the uncertainty distribution as a Gaussian.

Consider the three-dimensional volume of parameter space for our three neutrino masses: $m_1$, $m_2$, and $m_3$. The two pieces of data above resemble two thin slices cutting through our volume. The regions outside of these slices are excluded. The intersection of these two slices then forms two filaments, one falling on each side of the $m_1= m_3$ plane. Each filament therefore represents the permissible regimes of what are known as the normal and inverted hierarchies.   We shall seek to evaluate the posterior probability that a given filament contains the true set of masses. Before we can achieve this, we must first establish a consistent set of priors.
   
\subsection{Three-dimensional Priors}  
\label{sec:simplelogprior}
Our prior on the three neutrino masses, $ \pi(m_1, m_2, m_3)$, ought to reflect our state of belief before the data arrived. The conventional approach would be to assign an uninformative prior, and this corresponds to a uniform prior in each $\log m$.  Care must be taken in monitoring the influence of the breadth of this prior, since in practice some hard bounds must be imposed.   There are cases where the adoption of a linear prior may be justifiable, but this involves introducing information based on an understanding of the underlying distribution. For  example, for the case of the cosmological constant (see e.g. \cite{1987PhRvL..59.2607W}), a linear prior can be introduced if one is willing to assume that the bulk of the distribution $p(\Lambda)$ lies far above the value under consideration. Whether an equivalent assumption can be justified for the case of neutrino masses remains an open question. 

An illustration of the scenario we face is given in Figure \ref{fig:m1m3}.  The red dashed contours show the region allowed by oscillation data. They represent the two filaments discussed above, but now in projection. Their widths vary considerably since we are in the logarithmic domain (for visual clarity we use a broad $10 \, \sigma$ contour). The complementarity of the cosmological data is highlighted by the solid black curve, which provides an upper bound on the sum of the three masses  $\sumnu<0.12$ eV, thereby truncating the extent of the contours at larger masses. Note that due to freedom in the value of $m_2$,  the sum $\sumnu$ is not uniquely defined in the $m_1-m_3$ plane. What is of interest here is a contour of the surface $m_1 + m_2 + m_3 = 0.12$ eV which intersects the filaments. To create this we use the measurement of $\Delta m_{12}^2$ from equation (\ref{eq:m12}) to specify $m_2$. From this perspective it becomes clear that even with only the oscillation data at hand, before looking at \emph{any} cosmological data, we should consider it reasonably likely that the sum of the masses is of the order 0.1 eV (i.e., towards the lower limit allowed by oscillation data). This is because towards very large masses, the uncertainty in the measurement corresponds to a vanishingly small interval in $\log m$. Therefore our posterior on the sum of the masses, $\sumnu$, is heavily penalised towards large masses.  In stark contrast,  if we had adopted an informative uniform prior  (i.e., a uniform prior in $m$ rather  than $\log m$), we would have been repeatedly surprised as a sequence of modern cosmological surveys failed to unearth the influence of the neutrino masses. 

The crucial factor here is the influence of the third mass parameter $m_2$, which has been marginalised out in Figure \ref{fig:m1m3}. The filaments vary in width in $\log m_2$, in much the same manner as can be seen in the other two dimensions. This evolution in depth,  $\Delta \log m_2$, is  visually represented by the colour of the shaded areas.   In the regime of the normal hierarchy, and for the case of a vanishingly small $m_1$,  the range of permissible $m_2$  values at any given $(m_1, m_3)$ pair is governed by our uncertainty in the quantity $\Delta m_{12}^2$, of approximately $1.8 \times 10^{-6} \, \mathrm{eV}^2$.   Once we reach very low masses,  $m_1 < 10^{-3}$ eV, then we settle at a minimum value of $m_2 \sim 0.0087$ eV. Thus the space available within the $\log m_2$  domain saturates at approximately $0.022$. Let's compare this with the available parameter space in the inverted hierarchy. In this case, as we take the limit towards one of the neutrinos being massless, then we are left with $m_2 \sim 0.05$ eV. This corresponds to a much narrower range, approximately  $\Delta \log m_2 \sim 6 \times 10^{-4}$. Thus the sheer volume of parameter space that our data has been able to exclude is highly asymmetric with respect to the $m_1 = m_3$ plane.   

Our posterior belief in the two hypotheses, the inverted hierarchy and the normal hierarchy, relates to the integral of the posterior distribution over their respective areas. Provided our prior is sufficiently broad, each integral is dominated by the low mass regime, where the contours are at maximal thickness. Due to the saturation of $\Delta \log m_2$,  our result becomes completely insensitive to the breadth of our prior. The evidence ratio is well approximated by simply  $\Delta_{\text{NH}} \log m_2  / \Delta_{\text{IH}} \log m_2 \simeq 32$. Thus we ought to favour the normal hierarchy over the inverted hierarchy with odds of $32:1$. This will strengthen further if cosmological bounds on $\sumnu$ are imposed, as they  penalise the inverted scenario more aggressively than the normal hierarchy, as Fig.~\ref{fig:m1m3} illustrates.

We are drawn to this conclusion because the oscillation data have eliminated a highly asymmetric region of parameter space in the logarithmic domain.  This may not seem particularly intuitive, so it is instructive to consider similar scenarios where our posterior belief is driven by the volume of the available parameter space.  One example mentioned earlier is the identifying the orientation of a concealed stick - this is straightforward to verify numerically.  Another analogy which may be helpful to consider is the abundance of measurements  whose numerical values start with a ``1" or a ``9".  An intuitive response might be that they arise with equal frequency, but the reality is that there is almost a seven-fold difference in their respective abundances  (in any field of science)\footnote{This is known as Benford's law, which applies to   many real-life sets of numerical data \cite{Benford}.}. The most straightforward interpretation of this phenomenon is, again, due to the availability of parameter space in the logarithmic domain.  

\subsection{Mass Degeneracy}

It is interesting to consider a hypothetical special case: What would happen if $\Delta m_{21}^2$ had been perfectly measured to be identically zero?  This approximation is often used in cosmology since the solar mass splitting has  virtually no observational consequences for cosmological data.  Then there is no third dimension to worry about in Figure 1. The value of $m_2$ is defined at each point in the plane, so the shaded area would all be the same colour. And we therefore no longer have any asymmetry between the two hierarchies if we neglect any cosmological information. Our conclusion within this hypothetical scenario would be that both hierarchies are equally probable. In other words,  when adopting a logarithmic prior on the mass and setting the solar mass splitting to zero would make the two hierarchies equally likely. Adding a  cosmological bound breaks the symmetry and, in particular shows a preference for NH with odds 1.8:1 when $\Sigma<0.12$. This highlights the importance of performing the model selection analysis with the correct dimensionality in parameters and prior space. It is the smallness of the solar mass splitting (yet the fact that it is non zero)  relative to the atmospheric splitting that makes the Occam factor  favour the normal hierarchy using this particular prior.

\section{Hierarchical model}

In the previous section we presented a pedagogical calculation illustrating how the Occam factor strongly favours the normal hierarchy. However the framework was a little too simplistic to allow robust conclusions to be drawn.    

Rather than restrict ourselves to a single pre-determined prior, of fixed position and breadth, it is advantageous to work with a \emph{hyperprior}.  In practice we introduce an entire family of priors and effectively marginalises over them.  Hyperpriors are more flexible and make the posterior less sensitive to the  prior itself.    We shall operate in a five-dimensional parameter space, consisting of the three neutrino masses (for brevity we shall denote the mass vector  $\mathbf{m}$),  and two \emph{hyperparameters} $\mu$ and $\sigma$ (see e.g.,\cite{tiao1973some, mackay1996hyperparameters, hobson2002combining}), as follows
\begin{equation}
 \label{eq:hyperprior}
p( D, \mathbf{m},  \mu, \sigma | \mathcal{H })  = p( D |  \mathbf{m},  \mathcal{H })  p(  \mathbf{m} |  \mu, \sigma,  \mathcal{H })   \pi( \mu,  \sigma | \mathcal{H })  \, .
\end{equation}
Here $D$ represents  the data vector and ${\cal H}$  the hypothesis i.e., normal or inverted hierarchy, and  $\pi( \mu,  \sigma | \mathcal{H })$ is the hyperprior. 

To reiterate, the prior ought to reflect our state of belief before the data arrived. The three masses are indistinguishable before the data arrives, so our prior \emph{must} reflect this symmetry. This feature is also known as exchangeability. We note that this condition is satisfied if the three masses share a common origin, a hypothesis that is adopted in many models for the neutrino masses. 
 So we shall consider that three masses, $m_i$, $m_j$, $m_k$ are drawn from a common prior.   Then, based on their relative sizes, these three masses are assigned to $m_1$, $m_2$ and $m_3$, as defined earlier.  In order to specify the position and breadth of the prior, we shall introduce the variables $\mu$ and $\sigma$ which control the mean and standard deviation respectively. This specifies the prior on each mass \emph{before any ordering takes place}.  Thus the prior propagated through to  $m_1$ will tend to prefer lighter masses values than the effective prior on $m_3$.
Further we adopt a prior of the form

\begin{equation}
\label{eq:mu}
p( \log  \mathbf{m}  | \mu, \sigma) \sim {\cal N}(\mu, \sigma) \,,
\end{equation}
 where ${\cal N}$ denotes the normal, Gaussian, distribution.  For the purposes of performing Bayesian model selection, the quantity which we seek to evaluate is the evidence, $p( D | \mathcal{H })$. The inference procedure may then be decomposed into the following tiers, as is characteristic of Bayesian hierarchical models.  

\begin{equation}
 \label{eq:bayes}
p( D | \mathcal{H })  =  \iint p( D |    \mu, \sigma,  \mathcal{H })  \pi( \mu,  \sigma)  d \sigma d \mu  \, ,
\end{equation}

 \begin{equation}
 \label{eq:hyperbayes}
p( D |    \mu, \sigma,  \mathcal{H }) =  \int p( D |  \mathbf{m}, \mathcal{H })  p(  \mathbf{m} |  \mu, \sigma,  \mathcal{H })    d \mathbf{m}  \, .
\end{equation}
  
Note that our analysis differs significantly from that of Ref.~\cite{2016Hannestad}. Although not stated explicitly, they appear to make the following identification for the case of the normal hierarchy (where $m_1$ takes on the role of the lightest neutrino), 

\begin{equation}
p(D | m_1) \pi(m_1) = \iint  p(D |  \mathbf{m}) \pi(  \mathbf{m})  \delta(m_2^2 - m_1^2 - \Delta m^2_{12})  \delta(m_3^2 - m_1^2 - \Delta m^2_{13})   dm_2 dm_3\,;
\end{equation}
$\delta$ denotes the Dirac delta function.  The relation above implicitly assumes that \emph{informative} uniform priors are imposed upon both $m_2$ and $m_3$. This disproportionately favours masses with a value close to an imposed upper bound.  To illustrate the potential issues which can arise with informative priors, imagine applying the same technique to other sets of particles within the Standard Model. How should we proceed if we did not know the absolute values of any quark or lepton masses, but were given approximate values of their mass splittings and the sums of their masses. In each case, our estimation of the up, down, and electron masses would be heavily biased, since they lie several orders of magnitude below their heavier counterparts. 
Likewise, in the current circumstance, the lightest neutrino could be several orders of magnitude below $0.1$ eV. Theoretical considerations set the minimum viable neutrino mass at approximately $10^{-13}$ eV \citep{davidson2007smallest}. 
A different approach, which we do not explore here, is to  introduce a discrete hyperparameter  which assigns equal odds to the two hierarchies given the mass splitting obtained from oscillations constraints (an a posteriori decision, rather than a fundamental prior). This is for example discussed in \cite{referee_request} where the uncertainty on the mass splitting is neglected and a linear, improper, prior on the mass of the lightest neutrino is adopted.

 \begin{figure}
 \centering
\includegraphics[width=75mm]{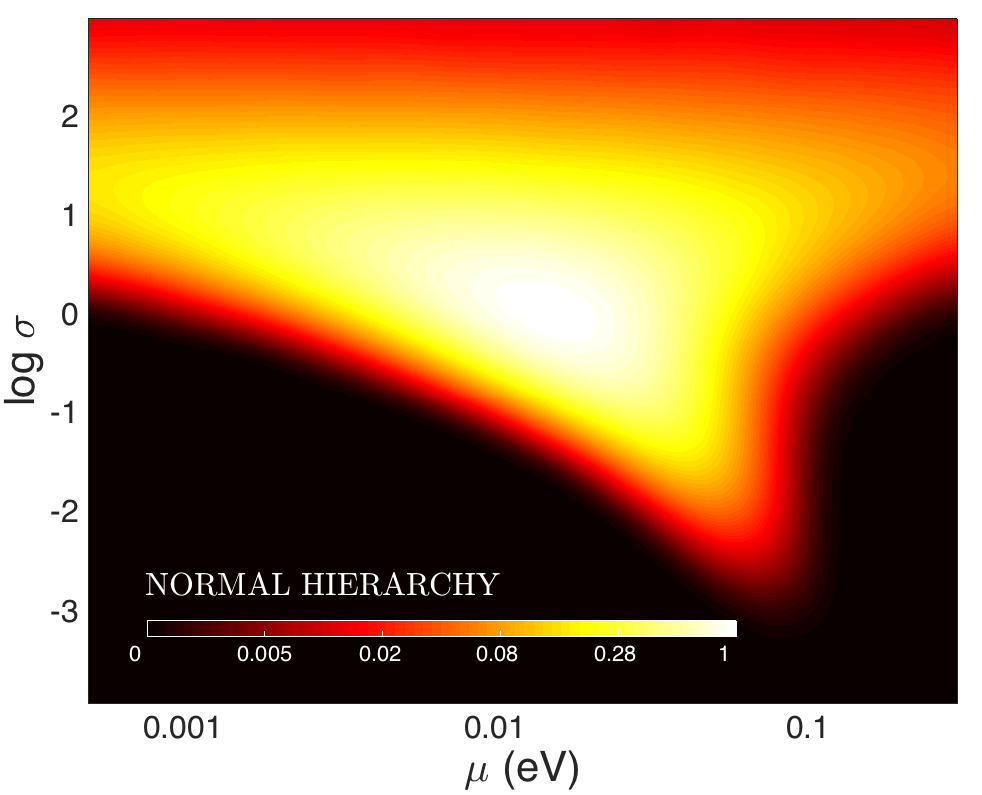}
\includegraphics[width=75mm]{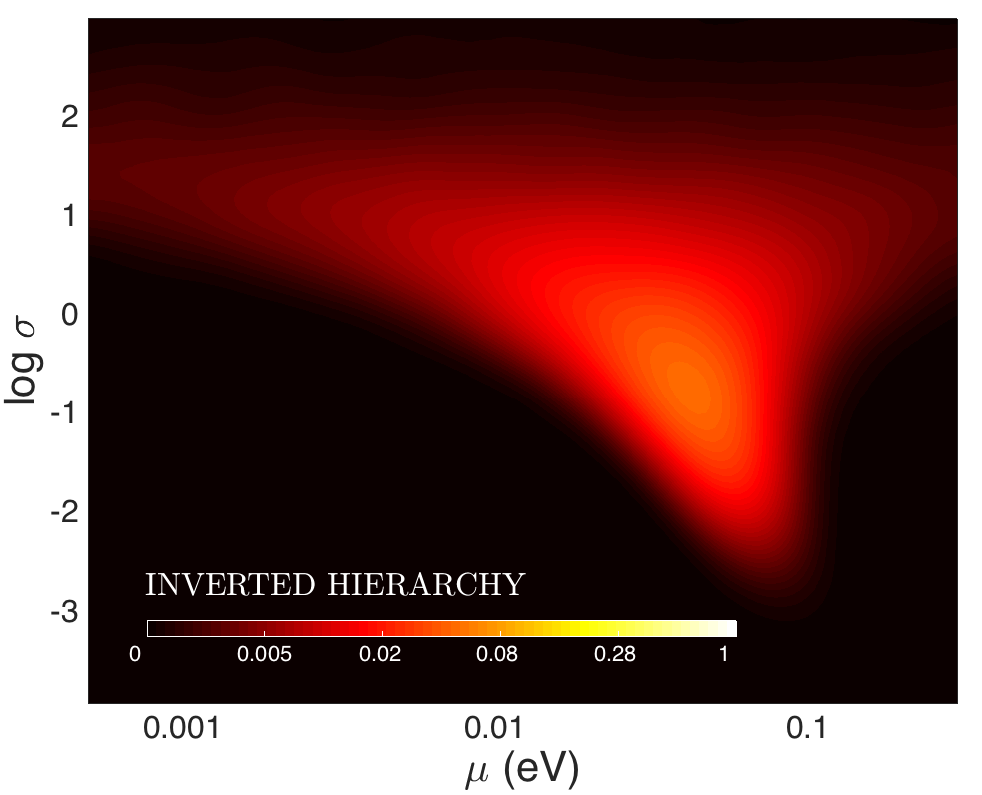}
\caption{Marginal likelihoods of the hyperparameters $\mu$ and $\sigma$, as defined in equation (\ref{eq:hyperbayes}). In each panel the same  data is being used, namely the two measurements of the mass splittings as given in equations (\ref{eq:m13}) and  (\ref{eq:m12}), and an upper bound is imposed on the sum of the neutrino masses $\sumnu < 0.15$ (95\% C.L.). The left hand panel corresponds to the case of a normal neutrino mass hierarchy, while the right hand panel reflects the inverted ordering scheme.  Note that a logarithmic colour scheme is employed in order to enhance the visibility of the right hand panel.  The ratio of the two integrated regions yields a Bayes factor, $K$, of  31. Results for other values of $\sumnu$ are presented in Table \ref{tab:evidence}.} \label{fig:musigma}
\end{figure}

\subsection{Methodology}

Here we briefly outline the procedure we use to numerically evaluate the likelihood of the hyperparameters $p( D |   \mu, \sigma, \mathcal{H })$. 

\begin{enumerate}

\item For a given value of $\mu$ and $\sigma$, this defines a probability distribution from which we draw $N = 10^7$ sets of three samples. For each set of three, the masses are ranked from smallest to largest, here we shall designate these ranks as $S$, $M$, and $L$. 

\item For each set we compute the following four quantities: $L^2 - M^2$; $L^2 - S^2$;   $M^2 - S^2$; and $\Sigma \equiv S+M+L $.

\item In order to determine the likelihood of the two hyperparameters, we  compute the inner integral of equation $(\ref{eq:bayes})$. For the case of the normal hierarchy  this becomes
\begin{equation}
 \eqalign{
p(D | \mu, \sigma, \mathcal{H}_{NH}) &=  \int  p( D |  \mathbf{m}, \mathcal{H}_{NH})  p(\mathbf{m} |  \mu, \sigma,  \mathcal{H}_{NH})  d \mathbf{m} \cr \, 
& \simeq   \frac{1}{N} \sum_{i=1}^N p(D_{L^2 - M^2} | \mathbf{m}_i) p(D_{M^2 - S^2} |  \mathbf{m}_i ) p( D_\Sigma | \mathbf{m}_i)\,.
 }
\end{equation}
 Here $\mathbf{m}_i$ denotes the mass vector of the $i$th random sample, while the $D_x$ terms represent the different components of the data vector. For example, $D_\Sigma$ represents the measurement of $\sumnu$ as derived from cosmological density perturbations.

\item Repeat Step 3 for the case of the inverted hierarchy. This involves a reinterpretation of the squared mass splittings as follows

\[
p(D | \mu, \sigma, \mathcal{H}_{IH}) \simeq \frac{1}{N} \sum_i p(D_{L^2 - S^2} | \mathbf{m}_i) p(D_{L^2 - M^2} |  \mathbf{m}_i ) p( D_\Sigma | \mathbf{m}_i)\,. 
\]

\item Repeat Steps 1-4 for 100 x 100 different values of $\mu$ and $\sigma$, across a logarithmically spaced grid. This yields the joint probability distribution shown in Figure \ref{fig:musigma}. The prior is uniform across this region, therefore all of the visible structure has been imposed by the dataset. 

\end{enumerate}

In order to test our numerical precision, we repeated this procedure with a smaller number of samples,   $N = 10^6$. This was found to modify the resultant Bayes factors by less than $1\%$.

 \section{Results}
 
Figure \ref{fig:musigma} illustrates the evidence for the hyperparameters, $p(D | \mu, \sigma)$.  The left hand panel corresponds to the case of the normal hierarchy.  Low values of $\sigma$ are disfavoured by the requirement of a significant mass splitting. Meanwhile very large values of $\sigma$ are penalised because they are less predictive. Cosmological information provides the third boundary, eliminating the largest values of $\mu$. Nonetheless a substantial area of parameter space retains a high  likelihood. 
By contrast, in the right hand panel we see the region of parameter space associated with the inverted hierarchy. For a linear colour scale the features are barely visible, so here we introduce a logarithmic scale.  The  likelihood suffers both in terms of its peak value (the maximum   is a factor of fifteen lower than  in the normal hierarchy) and in terms of its confinement to a more restricted region of parameter space. 

\begin{figure}
 \centering
\includegraphics[width=75mm]{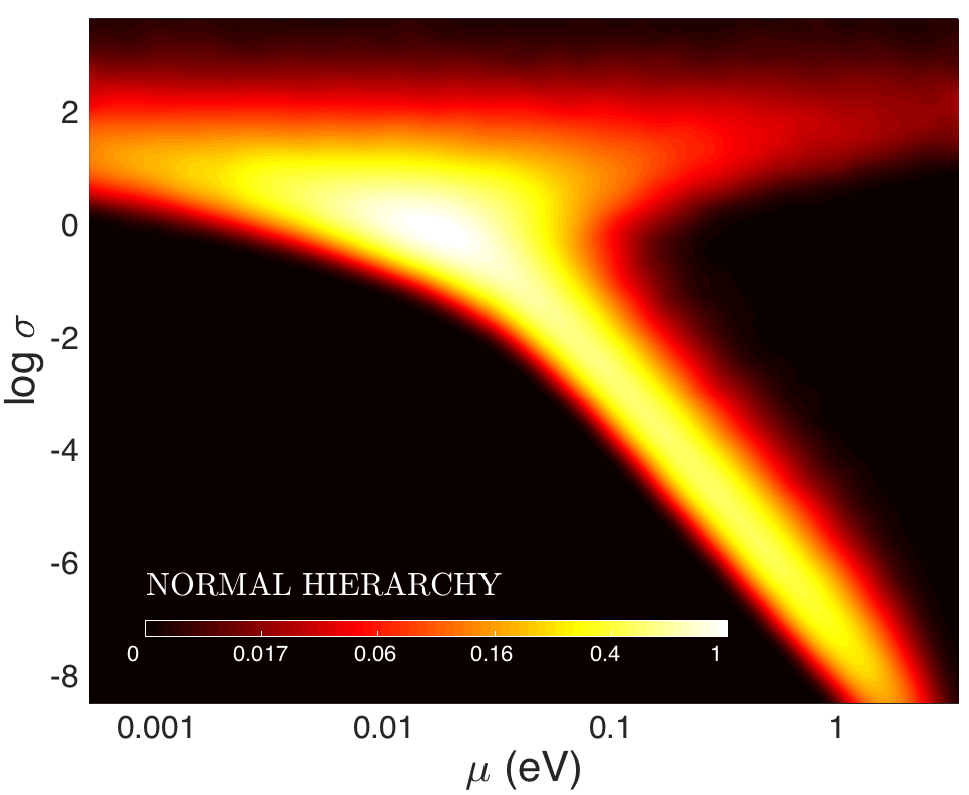}
\includegraphics[width=75mm]{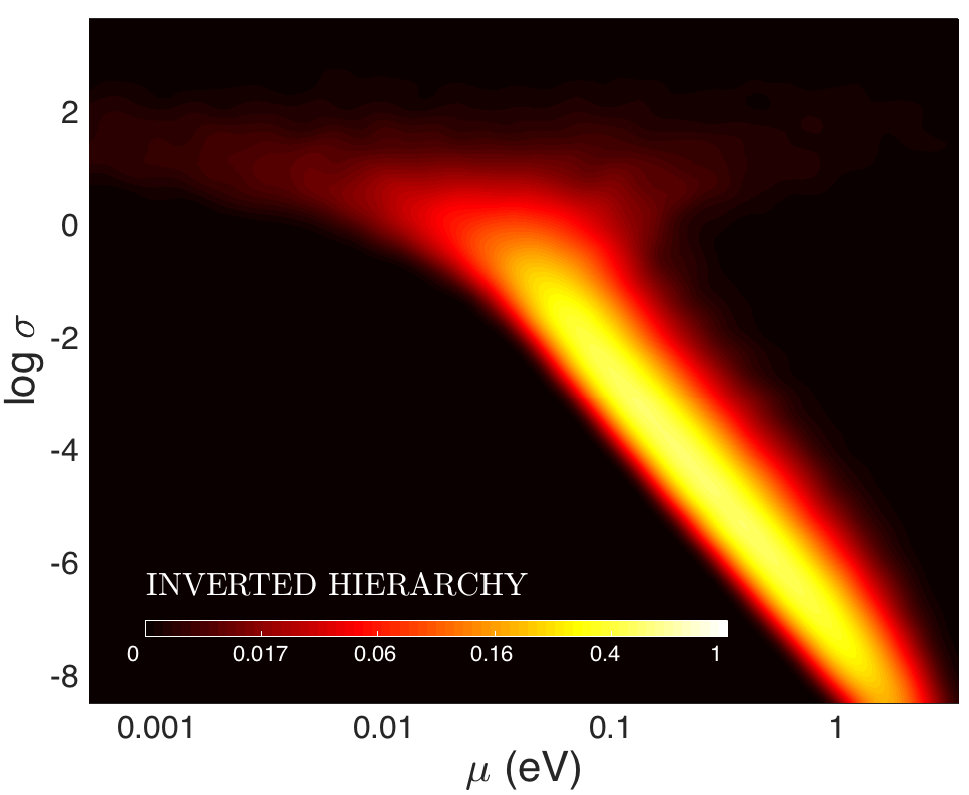}
\caption{As Fig.~\ref{fig:musigma} but  without the cosmological upper bound imposed on the sum of the neutrino masses. Instead we impose the laboratory constraint $\sumnu < 6.9\,$eV, yielding odds of $2.6:1$. Comparison with Fig.~\ref{fig:musigma} shows how cosmology has disproportionately eliminated the space available for the inverted hierarchy.}\label{fig:nocosmo}
\end{figure}

Another striking feature of Figure  \ref{fig:musigma} is that ratio between the two panels, at any given point in the $\mu-\sigma$ plane, favours the NH.  So we should anticipate that our results are not particularly sensitive to our choice of hyperprior, $\pi(\mu, \sigma)$. 
For comparison, in Fig.~\ref{fig:nocosmo}  we show the corresponding distribution without imposing the cosmological bound.  The comparison  of Figs.~\ref{fig:musigma} and \ref{fig:nocosmo}, clearly illustrates  how cosmology has wiped out the space available for the inverted hierarchy.

\begin{figure}
 \centering
\includegraphics[width=85mm]{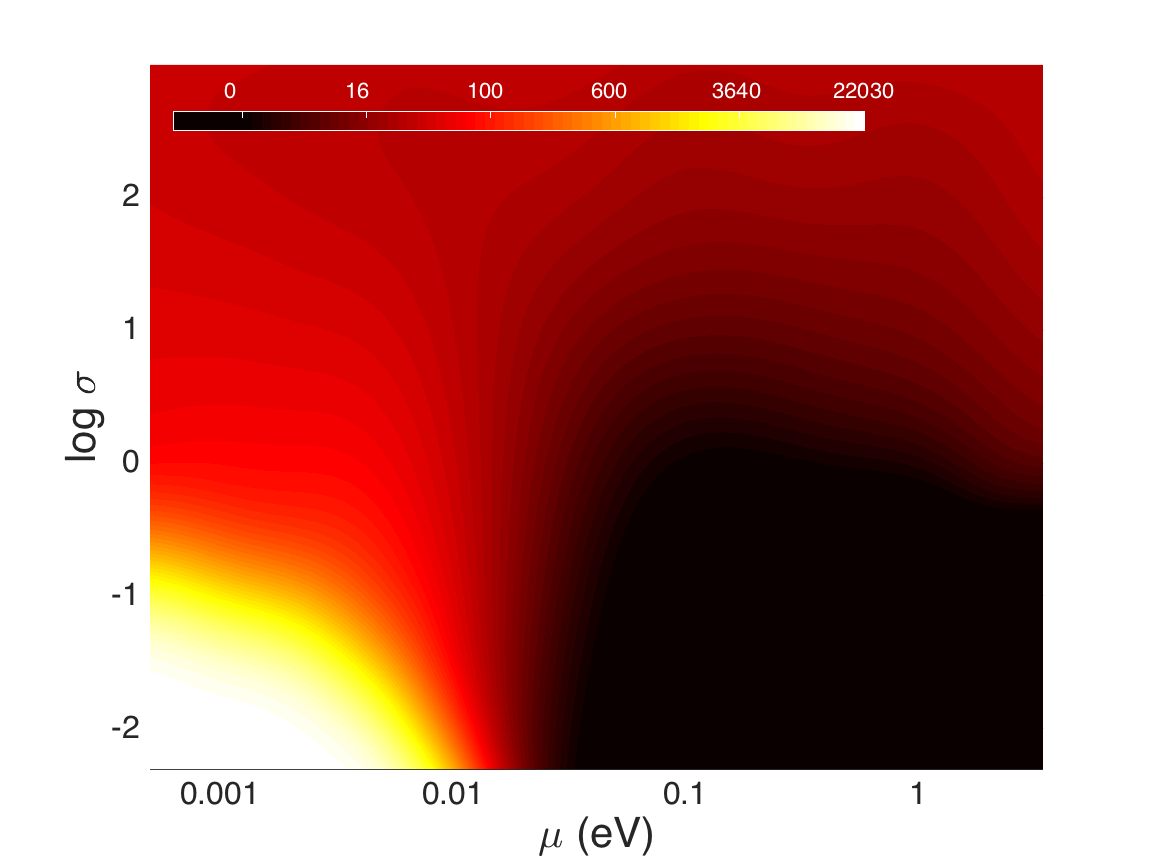}
\caption{An illustration of how the evidence ratio evolves across the $\mu - \sigma$ plane, due to data from neutrino oscillations. Notice that for broader distributions,  $\log \sigma>0$, we consistently find evidence values of $K \simeq 30$. This limiting case is analogous to that of our earlier investigation in Section \ref{sec:simplelogprior}, since a large value of $\sigma$ ensures that the lightest mass is likely to be very much smaller than the other two. }\label{fig:evidence}
\end{figure}

\subsection{Bayesian Evidence}

Our posterior belief in the two hypotheses, the normal and inverted orderings, is given by
\begin{equation}
 \frac{p(\mathcal{H}_{NH} | D)}{p(\mathcal{H}_{IH} | D)}  =  \frac{p( \mathcal{H}_{NH})}{p( \mathcal{H}_{IH})}   \frac{p(D| \mathcal{H}_{NH})}{p(D| \mathcal{H}_{IH})} \,.
\end{equation}
In principle we are free to introduce a bias in the ratio $p( \mathcal{H}_{NH})/p( \mathcal{H}_{IH}) $, but here we consider the prior probabilities of the two models to be equal. The important quantity in influencing the strength of our posterior belief is therefore the evidence ratio, $p(D| \mathcal{H}_{NH})/p(D|\mathcal{H}_{IH})$, also known as the Bayes factor $K$ defined in equation (\ref{eq:bayesfactor}). This quantity is evaluated using equation (\ref{eq:bayes}), while again employing an uninformative prior $\pi(\mu, \sigma)$, as given by

\begin{equation}
\eqalign{
p(\log \mu) &\sim U(\mu_{min}, \mu_{max})   \cr
p(\log \sigma) &\sim U(\sigma_{min}, \sigma_{max})
}
\end{equation}
where $U$ denotes the uniform distribution.  For computational purposes we set hard bounds of $[\mu_{min} = 10^{-4} eV, \mu_{max} = 0.5 eV]$ and $[\sigma_{min} 0.005= , \sigma_{max} = 20]$, broadening either of these limits has no impact on our results. Provided an upper bound is placed on the sum of masses, in addition to the mass splittings, then the data is capable of restricting the marginal likelihood associated with the hyperpriors to a finite range.
We begin by noting that in the absence of neutrino oscillation data, i.e, setting the
uncertainties of the mass splittings to be infinite, then our priors yield odds of the two hierarchies of precisely 1 : 1. In fact this is a generic feature of \emph{any} prior that is invariant to permutations of its elements, i.e. $p(m_1, m_2, m_3) = p(m_3, m_1, m_2)$, as this ensures $p(m_1 > m_3) = p(m_1 < m_3)$.

Figure \ref{fig:evidence} illustrates how the evidence ratio changes as we traverse the $\mu - \sigma$ plane. Here we make use of data from neutrino oscillations. Note that  for  broader distributions, $ \log \sigma > 0$, we consistently find evidence values of $K\sim 30$ is agreement and analogy with the findings of Sec. 3.2.

In Table \ref{tab:evidence} we present the Bayesian evidence for a range of different degrees of precision for the sum of the neutrino masses. Hereafter, for the classification of the evidence, we refer to the slightly modified Jeffreys' scale \cite{Jeffrey} as discussed in \cite{kassraftery}. 
When only utilising data from neutrino oscillations, there is no significant evidence for one hierarchy  over the other, being classified as ``Weak". This result is qualitatively different from that obtained in Sec. \ref{sec:simplelogprior} where a simple log prior was used and highlights the usefulness of introducing the hyperprior.
The current robust limit  $\sumnu< 0.13$ eV from \cite{2016Cuesta} already provides odds of approximately 42:1, which translate into strong evidence in favour the normal hierarchy. Note that, if we instead use the Ly-$\alpha$ limit, $\sumnu < 0.12$ eV,  the odds increase to 50:1, if we also include the latest measurement of the Hubble constant \cite{Riess2016}  the odds increase to 90:1. Thus in this approach, the cosmological bound on $\sumnu$ is critical in driving the evidence in favour of the normal hierarchy.

\subsection{Sensitivity Analysis}

\begin{figure}
 \centering
\includegraphics[width=75mm]{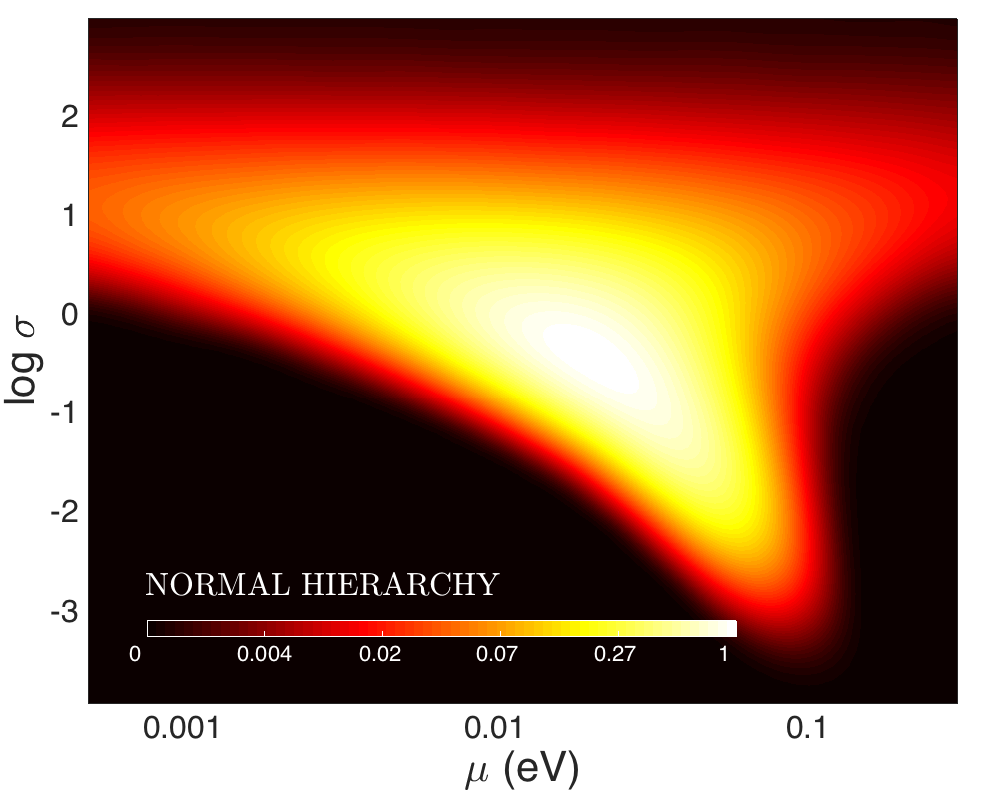}
\includegraphics[width=75mm]{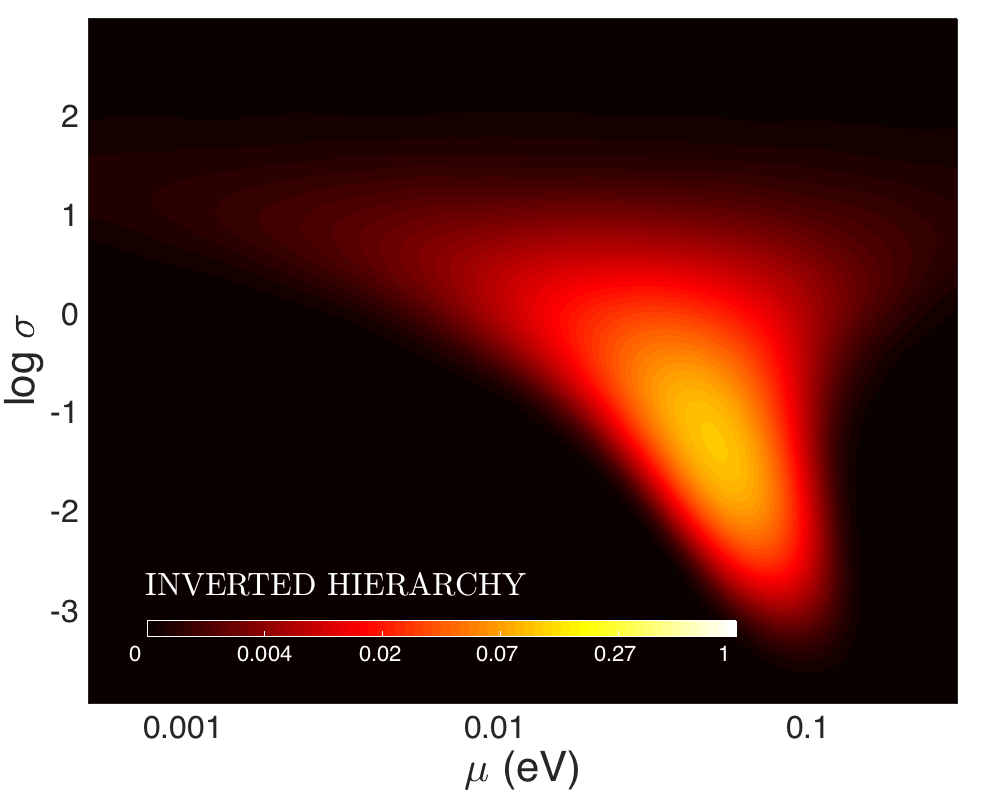}
\caption{Exploring our sensitivity to the choice of hyperprior. The above panels are in a similar format to Fig.~\ref{fig:musigma}, but here we show the posterior distribution on the hyperparameters  when imposing an informative prior of the form $\pi(\sigma) \propto \sigma^{-2}$. This enforces a bias in favour of closely clustered neutrino masses, which leads to a slight weakening of the evidence ratio, yielding odds of $17:1$ for $\sumnu < 0.15 \,$ eV (95\% C.L.), and $33:1$ for $\sumnu < 0.12 \,$ eV (95\% C.L.).}\label{fig:invmusigma}
\end{figure}

In order to explore our sensitivity to the choice of prior $\pi(\mu, \sigma)$, we shall now impose an informative prior on $\sigma$. Figure \ref{fig:invmusigma} shows the impact of adopting a hyperprior of the form $p(\sigma) \propto \sigma^{-2}$. This favours narrower distributions, which in turn slightly weakens the extent to which the NH is favoured over IH. For the illustrated example we set $\Sigma < 0.15 \,$eV, leaving us with odds of $17:1$. 
Meanwhile $\Sigma < 0.12\,$eV returns odds of $33:1$.

We conducted a further sensitivity test by replacing the normal distribution with a top hat distribution, described by the same set of parameters $\mu$ and $\sigma$. This was found to have a minimal impact on the evidence ratios.

\section{Discussion and conclusions}

By performing a likelihood analysis in its full three-dimensional parameter space, and adopting a minimally informative hierarchical model,  we demonstrate that the inverted hierarchy suffers from a much more extensive exclusion of its parameter space than the normal hierarchy. This can be largely attributed to the residual freedom in the mass of $m_2$ in the logarithmic domain. Central to this result is the cosmological bound on the sum of the neutrino masses. Without this information, the Bayesian evidence is weak, and does not significantly favour one ordering over the other, as it is apparent from Tab.~1 and 2.  As with all applications of Bayesian model selection,  there is some dependence on the choice of adopted prior. Nonetheless, we can draw some robust conclusions which have several broad implications.

\subsection{Implications in neutrino physics}

The determination of the nature of neutrinos, whether Majorana or Dirac fermions, is a major long-term unsolved problem in particle physics, with double beta decay experiments being the most sensitive probes to solve this riddle. These experiments could prove the existence of a rare double beta decay of nuclei without neutrino emission, a decay which is mediated by the internal exchange of the light Majorana neutrinos (other beyond the standard model physics may also contribute to the decay). In the minimal case, the expected half-lives of neutrino-less double beta decays  of nuclei depend on nuclear physics dynamics and on the neutrino mass scale and hierarchy. In fact, the neutrino mass hierarchy plays a key role in identifying the largest half-lives to be explored \cite{Murayama:2003ci}. In our opinion, the results derived in this work are very relevant to the development of next generation double beta decay future experiments. While it is true that, independently of the neutrino mass hierarchy, the discovery could be close to the upper bounds, for example 10$^{27}$ yr for Xe \cite{KamLAND-Zen:2016pfg}, the strong evidence for the normal neutrino mass hierarchy favours experimental techniques with potential to reach multi-ton active mass detectors and meet the requirements of very low background with, for example, topological signatures\cite{Granena:2009it,Chavarria:2016hxk}. 

Our results have also important implications for neutrino oscillation experiments and neutrino mass models. Neutrino mass hierarchy is among the physics goals of several proposed neutrino detectors \cite{Abe:2015zbg,Acciarri:2015uup,An:2015jdp,TheIceCube-Gen2:2016cap,Adrian-Martinez:2016fdl}. Experiments whose major scientific goal is the identification of the neutrino mass hierarchy are particularly affected by the conclusions of this paper. In particular, according to our findings, experiments with observables more sensitive to normal mass hierarchy are much more likely to yield a successful outcome.

 \begin{table*}  
\caption{Evidence ratios for the Normal and Inverted orderings and classification according to Jeffreys'  \cite{Jeffrey, kassraftery}, for a range of hypothetical cosmological data. Fig.~\ref{fig:musigma} illustrates the marginal  likelihood surface for the case $\sumnu < 0.15$ eV. Since our prior odds are equal, then the Bayes factor $K$ matches the quoted posterior odds.} 
\label{tab:evidence}
\begin{tabular}{ccccccccc} 
\hline
 $\sumnu$ (eV)    95\%            & $< 0.05$     &  $< 0.1$        & $<0.12$ & $<0.13$   & $<0.15$    &  $<0.2$ &  $<0.5$ &  $<6.9$ \\
   \hline                
Odds (NH/IH)      & 6500:1  & 95:1 &  53:1 & 42:1 &  31:1 & 18:1 & 7:1 & 2.6:1 \\ 
$\log K$                 & 9  & 4.5  & 4.0 & 3.7 & 3.4  & 2.9 & 1.9 & 1.0 \\ 
Classification      & Very Strong  & Strong  &  Strong & Strong & Strong & Positive & Positive & Weak \\ 
\hline
\end{tabular} 
\end{table*}

 \begin{table*}   
\caption{The same as Table  \ref{tab:evidence},  but now for the scenario where $p(D_\Sigma | \sumnu)$, the evidence for $\sumnu$ from cosmological data, peaks at $0.05 \,$eV. } 
 \label{tab:offset}  
\begin{tabular}{cccccccc} 
 \hline
 $\sumnu$ (eV)    95\%                   & $< 0.1$     &  $< 0.15$        & $<0.17$    & $<0.2$    &  $<0.25$  &  $<0.5$ &  $<6.9$\\
   \hline 
Odds (NH/IH)      & 225:1  & 33:1 &  24:1 & 18:1 &  13:1 & 6.3:1 & 2.6:1   \\ 
$\log K$                 & 5.4  & 3.5  & 3.2 & 2.9 & 2.5 & 1.8 & 1.0   \\ 
Classification      & Very Strong  & Strong  &  Strong &   Positive & Positive & Positive & Weak \\ 
\hline
\end{tabular} 
\end{table*}

\begin{figure}
 \centering
\includegraphics[width=120mm]{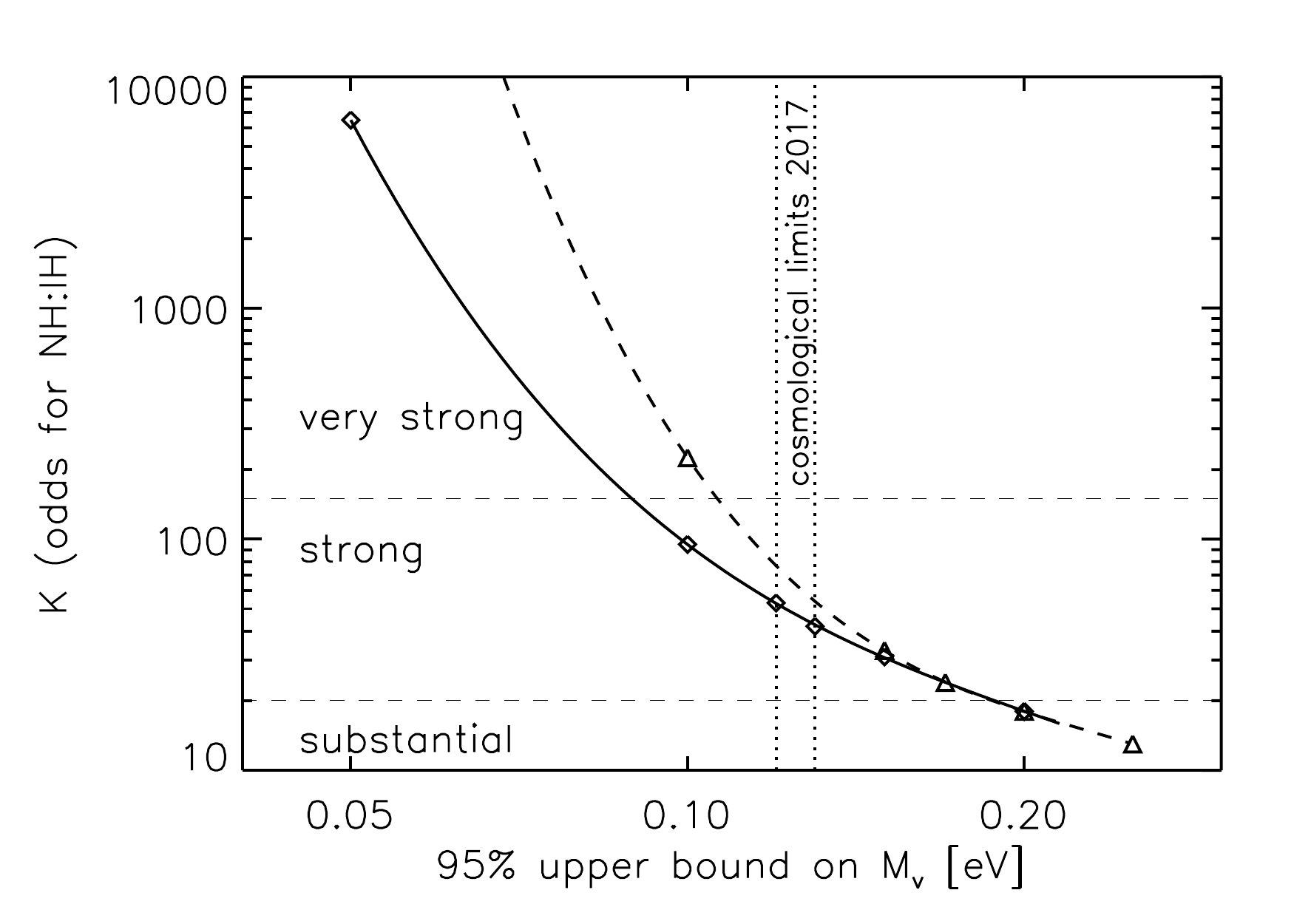}
\caption{Odds as a function of cosmological upper  limits  on the sum of neutrino masses $\sumnu$. The solid line corresponds to the (actual) case where the maximum of the  $\sumnu$ distribution is indistinguishable from zero. The dashed line correspond to a case where maximum of the  $\sumnu$ distribution is at $0.05$ eV.  The symbols correspond to the values reported in Tab.~\ref{tab:evidence} and \ref{tab:offset}.  Also indicated (vertical dotted lines) are the current limits from CMB and clustering  of galaxies \cite{2016Cuesta}  or Lyman $\alpha$ forest \cite{Palanque-Delabrouille:2015pga}. Jeffreys' interpretation of the Bayes factor values are also reported.  }   \label{fig:odds}
\end{figure}

\subsection{Interpreting results from upcoming surveys}

Fig.~\ref{fig:odds} serves as guidance for the interpretation of forthcoming results from cosmology. In order to facilitate a comparison with conventional cosmological analyses, we present our results in terms of the $95\%$ confidence bounds which would be achieved when imposing a uniform prior on $\sumnu$, and without utilising external information from neutrino oscillation experiments. The solid line corresponds to the family of Gaussian likelihoods $p(D_\Sigma | \sumnu)$, of varying standard deviation but each with maxima located at $0\,$eV, as motivated by the findings of recent analyses. Meanwhile the dashed line demonstrates the impact of shifting the maximum likelihood value to $0.05\,$eV.  These lines illustrate the degree of belief we would reach in the normal neutrino hierarchy, when this hypothetical cosmological data is combined with the squared mass splittings presented in Section 3.1. They were computed by interpolating the values of Tab.~\ref{tab:evidence}, shown as symbols. What if these hypothetical future measurements provide instead a likelihood for $\sumnu$ which is centred at a value close to the current 95\% upper limit? In this case, if this value is e.g., $0.1\,$eV, the preference for normal hierarchy is much weaker but still a preference remains. For example for 1-$\sigma$ errors of $\{0.005, 0.01, 0.025, 0.05\}\,$eV corresponding to 95\%  upper limits of $\{0.11,0.12,0.15, 0.2\}\,$eV, one would obtain odds of $\{3.7, 5.3,11, 12\}$ to 1 in favour of NH.

\begin{figure}
 \centering
\includegraphics[width=120mm]{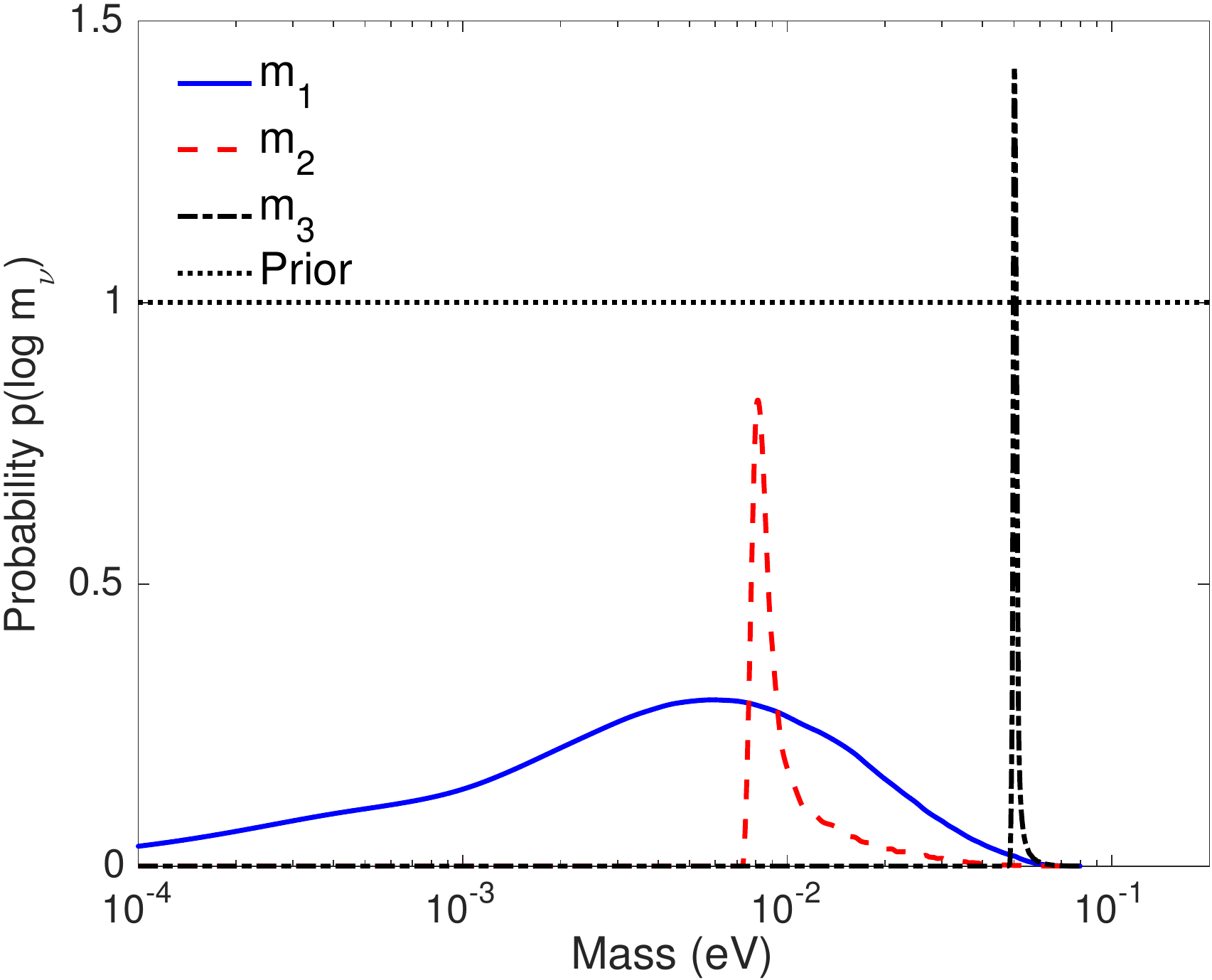}
\caption{The posterior distribution on the three individual masses, for the case of the normal hierarchy. To reach these constraints we utilise   $\sumnu < 0.12\,$eV cosmological bound. The amplitudes of the distributions are rescaled for clarity. } \label{fig:masses}
\end{figure}

\subsection{Constraints on the individual masses}

Figure \ref{fig:masses} presents the posterior distributions on the three individual neutrino masses. In light of our findings in the previous section, here we only consider the normal hierarchy's contribution to the posterior. Unsurprisingly the lightest mass is the least well constrained, with a mild preference that it is not very far in log space from the other two masses. Meanwhile the two heavier masses, $m_2$ and $m_3$, possess sharper but highly asymmetric posteriors.  From these posterior distributions we may derive medians and credible intervals for each mass eigenstate, yielding $m_1 = 3.80^{+26.2}_{-3.73} \, \text{meV}; m_2 =  8.8^{+18}_{-1.2} \,  \text{meV}; m_3 =  50.4^{+5.8}_{-1.2} \,  \text{meV}$ ($95\%$ CL). %

Finally, the dotted horizontal line illustrates our belief in each of the masses before the data arrives. We stress that the difference between the prior and the posterior distribution is unequivocally driven by the data. 

\subsection{Constraints on the sum of the masses}

\begin{figure}
 \centering
\includegraphics[width=120mm]{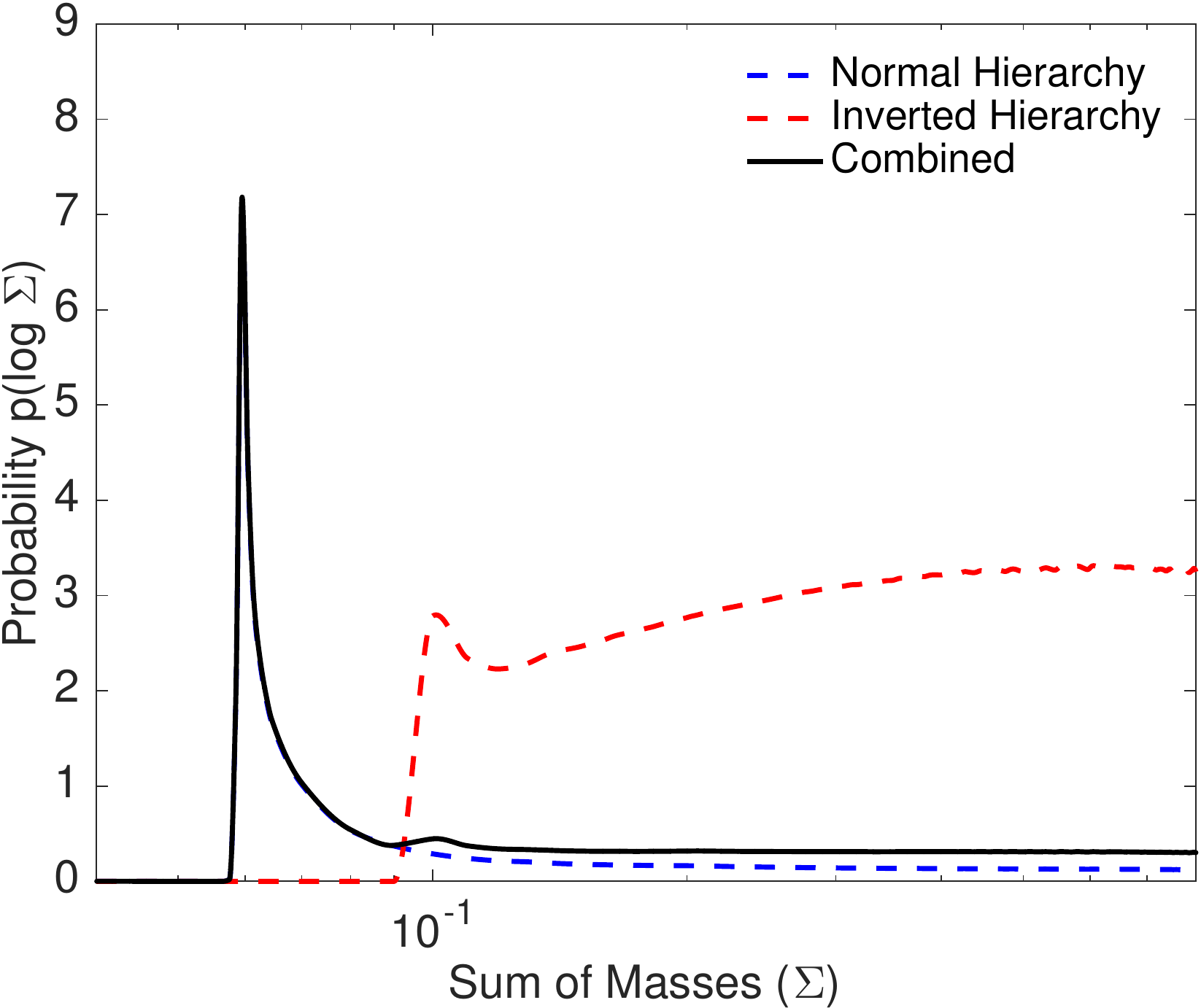}
\caption{The posterior distribution on the sum of the three neutrino masses. No  cosmological information is included here, instead we utilise $\sumnu < 6.9\,$eV. Normalisations are rescaled for clarity. } \label{fig:sumofmasses}
\end{figure}

In order to perform a rigorous inference of cosmological parameters,  by exploiting data from cosmological surveys, it is necessary to ascribe a prior on the sum of the neutrino masses $\sumnu$. For example, the Planck team either assign a fixed fiducial value of $\sumnu = 0.06\,$eV, or explore a uniform prior. However, note that even a uniform prior on the three individual masses would not yield a uniform prior on the sum.  Here we advocate that the posterior distribution from the oscillation experiments should form the basis of our prior for cosmological surveys. In Figure \ref{fig:sumofmasses} we illustrate this distribution, given as a weighted sum of the two components, specifically

\begin{equation}
p(\sumnu | D) = p_{\text{NH}}(\sumnu | D) p(\text{NH} | D)  + p_ \text{IH}(\sumnu | D) p(\text{IH} | D)   \, .
\end{equation}
The red and blue dashed lines illustrate  $p_{\text{NH}}(\sumnu | D)$ and $p_ \text{IH}(\sumnu | D) $ respectively, while the solid line depicts their weighted combination $p(\sumnu | D)$. Here our data vector  $D$ is comprised of the two measured mass splittings, and the bound on $\sumnu$ derived from laboratory experiments (i.e., no cosmological information). This leads an odds ratio between the two hierarchies of $2.6:1$. Since $p(\sumnu | D)$ is potentially useful for future analyses of cosmological data, it has been made publicly available\footnote{At the following URL: \url{https://github.com/frgsimpson/neutrino-priors}}.

By introducing additional cosmological information, the sum of the masses can also be resolved,  Our fiducial Gaussian bound of $\sumnu <0.13 \,$eV yields $\sumnu =  64^{+61}_{-5} \,$meV. The mode of the posterior distribution $p(\sumnu | D)$ lies significantly below the quoted median value, at $60\,$meV. As before, this posterior is composed of two distinct contributions from the normal and inverted hierarchies. The measurement does not drastically change in the case of the normal hierarchy:  $\sumnu =  63^{+50}_{-4.7} \,$meV. Meanwhile the inverted hierarchy continues to prefer considerably larger values, $\sumnu =  112^{+69}_{-17} \,$meV  (all 95\% credible intervals). 

Figure \ref{fig:sumofmasses}  also provides us with another perspective on how the cosmological data's preference for the normal hierarchy arose. Once the mass splittings were measured, the inverted hierarchy much prefers larger values of $\sumnu$, and it is this regime which has now been largely excluded by cosmological data. 

\subsection{Conclusions}
 
 We have presented a joint analysis of data on the masses of the three neutrinos, using Bayesian hierarchical modelling. We find that the combination of laboratory and cosmological measurements already provide strong evidence in favour of a normal hierarchy. The complementarity of these datasets also permit an inference of the individual masses, $m_1 = 3.80^{+26.2}_{-3.73} \, \text{meV}; m_2 =  8.8^{+18}_{-1.2} \,  \text{meV}; m_3 =  50.4^{+5.8}_{-1.2} \,  \text{meV}$, and their sum   $\sumnu =   64^{+61}_{-5} \,$meV ($95\%$ credible intervals).

The extent to which our belief in the inverted hierarchy diminishes is related to the  collapse in its available parameter space. Within the context of the inverted hierarchy, only contrived combinations of neutrino masses remain compatible with the data.  While this effect has been known to exist in a one-dimensional context, here we have demonstrated that the effect is significantly amplified when accounting for the full three-dimensional domain.
Our results differ (quantitatively, not qualitatively) from those of Refs.~\cite{2016Hannestad,Mena17} who use an informative uniform prior and find that the odds for NH:IH are about 2:1 or 3:1, only, depending on the dataset combination. Such a prior tends to favour high masses, prohibiting the masses from possessing significant variance in the logarithmic domain. When dealing with parameters whose uncertainty spans several orders of magnitude, such as in this case, we argue that a less informative prior should be adopted.

Due to the differences in parameter space, the first significant digit of a given neutrino mass is much more likely to be a `1' than  a `9', in accordance with Benford's law. This may at first seem unintuitive, yet it is incontrovertible. One does not require knowledge of the formation mechanism in order to make this statement.  In a similar vein, the mass hierarchy is very much more likely to be normal rather than inverted. We also found a slight preference for the lightest mass to be smaller than 5 meV.  These results, light smallest mass and normal hierarchy, agree well with predictions derived by, among other models, minimal SO(10) grand unified models \cite{Buccella:2017jkx}.

Is there a means by which our conclusion can be evaded? One can always resort to invoking unknown systematics large enough to invalidate the observational data, or hoping that the $\Lambda$CDM cosmological model is wildly inaccurate, but these appear to be highly speculative escape routes. As always, our posterior belief is to some extent influenced by our choice of prior. In this work we employed a minimally informative prior, and demonstrated that modest modifications to the prior impart only a minor impact on the evidence ratios. One could construct a more contrived set of priors which strongly favour inverted orderings. Or equivalently, one could amplify the prior on the hypothesis itself $p(\mathcal{H}_{IH})$. However there appears no clear justification for either of these actions.
  \\

\textit{Acknowledgements.---} This work is supported by Generalitat Valencia Prometeo Grant II/2014/050, by the Spanish Grants FPA2011-29678 of MINECO and by PITN- GA-2011-289442-INVISIBLES.
Funding for this work was partially provided by the Spanish MINECO under projects AYA2014-58747-P and MDM-2014-0369 of ICCUB (Unidad de Excelencia `Mar{\'\i}a de Maeztu').


\end{document}